\documentclass[12pt,a4paper]{article}

\def\be{\begin{equation}}
\def\ee{\end{equation}}
\def\bea{\begin{eqnarray}}
\def\eea{\end{eqnarray}}
\def\ba{\begin{array}}
\def\ea{\end{array}}
\def\pa{\partial}

\def\nn{\nonumber}

\def\Tr{{\rm Tr}}

\def\ga{\gamma}
\def\Ga{\Gamma}
\def\si{\sigma}

\def\de{\delta}
\def\ka{\kappa}
\def\al{\alpha}

\def\la{\lambda}
\def\La{\Lambda}
\def\eps{\epsilon}

\def\id{{\,\,{\rm l}\!\!\!1}}

\begin{document}

\begin{titlepage}
\begin{flushright}
IFT-20/2000\\  
hep-th/0009122  
\end{flushright}

\vspace{1cm}

\begin{center}
{\Large{\bf {Domain walls without cosmological constant
\\
in higher order gravity}}}
\end{center}

\vspace{1.5cm}

\centerline{\bf 
Krzysztof A. Meissner\footnote{e--mail: Krzysztof.Meissner@fuw.edu.pl} 
and Marek Olechowski\footnote{e--mail: Marek.Olechowski@fuw.edu.pl}
}

\vspace{0.5cm}
\centerline{\em Institute of Theoretical Physics}
\centerline{\em Warsaw University}
\centerline{\em Ho\.za 69, 00-681 Warsaw, Poland}

\thispagestyle{empty}

\vspace{1.5cm}
\begin{abstract}
We consider a class of higher order corrections with arbitrary power
$n$ of the curvature tensor to the standard gravity action in
arbitrary space--time dimension $D$. The corrections are in the form
of Euler densities and are unique at each $n$ and $D$. We present a
generating functional and an explicit form of the corresponding
conserved energy--momentum tensors. The case of conformally flat
metrics is discussed in detail. We show that this class of corrections
allows for domain wall solutions since, despite the presence of higher
powers of the curvature tensor, the singularity structure at the wall
is of the same type as in the standard gravity. However, models with
higher order corrections have larger set of domain wall solutions and
the existence of these solutions no longer depends on the presence of
cosmological constants. We find for example that the Randall--Sundrum
scenario can be realized without any need for bulk and/or brane
cosmological constant.   
\end{abstract}

\vfill

September 2000

\end{titlepage}

The idea of calculation of the effective 
quantum corrections to the energy--mo\-men\-tum tensors in 
the nontrivial gravitational background is relatively old 
\cite{BD,GMM}. If gravity is treated classically 
such a calculation corresponds to a one--loop result in 
quantum theory. If the background metric is conformally flat 
then there is no massless particle creation and the result 
for the vacuum polarization is local, therefore the use of the
effective lagrangian is justified.
These corrections are in the form of an expansion in 
powers of the curvature tensor.
They belong 
to two classes. One of them ("topological") consists of terms 
that can be obtained from an effective action in the first 
order formalism without need for any metric. The second class 
("non--topological") consists of terms coming from an action 
that necessarily involves a metric.

In this paper we are 
interested in the first class corresponding to the set of 
Euler densities of arbitrary order $n$ ($n$ being a power of the
curvature tensor) \cite{L}. 
When the dimension of the space--time $D$ is equal
to $2n$ the Euler density gives part of the conformal anomaly.
This type of anomaly is unique and
preserves scale invariance \cite{DS}, (the other type requires
introduction of a 
scale through regularization and vanishes for vanishing Weyl tensor). 
It is important to stress that these higher order gravitational terms
should always be included in the presence of the quantum matter fields 
-- for example in the case of the Standard Model
fields\cite{BD,GMM}.

There is another, independent motivation
to consider terms of higher order in the curvature tensor. Such terms
appear in the $\alpha'$ expansion of the string theory effective
action and it was shown that the quadratic terms can be put in the
form of the Gauss--Bonnet combination ($n=2$ Euler
density)\cite{strings}. 

It is interesting to note that Euler densities are the only higher
order
terms in the action that do not
introduce ghosts and assure ``Cauchy--like'' evolution of the initial
conditions.

{}From the form of the Euler densities it follows that for arbitrary
$D$ and $n$ all the energy--momentum tensors belonging to the first 
class do not have derivatives acting on the 
curvature tensor. 
For the case of the conformally flat metrics we find a generating 
functional for these energy--momentum tensors. 
We analyse the domain wall solutions in this class of models. 
We show that such solutions can be constructed since, even in the
presence of higher powers of the curvature tensor, the type of
singularity at the wall is the same as in the standard
gravity. In the usual domain wall metric the curvature has a
delta--like singularity on the wall. Thus, one would naively expect
that the higher powers of the curvature tensor introduce meaningless
expressions (higher powers of the Dirac delta function) which can not
correspond to any sources. 
The Euler densities seem to be the only combinations of the higher
order terms that do not lead to such meaningless expressions and
which allow for the domain wall metric with the usual sources.

The number of the domain wall solutions with higher order terms in the
action is larger  
than in the standard case and the cosmological constant is in general 
no longer crucial for their existence. It may even happen that the
Randall--Sundrum type scenario \cite{RS2} can be realized  
without bulk and/or brane cosmological constant. 
The fact that there exist brane--like solutions without any
cosmological
constant does not solve the cosmological constant problem -- it is, as
usual, the fine tuning problem
since the solution requires then some relations among the couplings.

In this paper we assume that all the higher order corrections to the
lagrangian are of the form of Euler densities which in $D$ dimensions
are defined (in the form notation) as 
\be
\mbox{\boldmath$I$}^{(n)}=\frac1{(D-2n)!}{\eps_{a_1a_2\cdots a_D}} 
R^{a_1a_2}\wedge \cdots R^{a_{2n-1}a_{2n}}
\wedge 
e^{a_{2n+1}}\wedge \cdots e^{a_D} 
\,.
\label{eq:I}
\ee
For $D=2n$ they are invariants and formally total derivatives.
However, careful regularization does not allow us to discard them
neither in the action nor in the equations of motion since they
correspond to the conformal anomaly.

If we have a lagrangian of the form 
\be
{\cal L}=-\sum_{n=0}^{n_{max}} \ka_n \mbox{\boldmath$I$}^{(n)}
\label{eq:L} 
\ee
then the equations of motion from the variation of the 
vielbein read (for $n \le (D-1)/2$) 
\be
\sum_n\frac{\ka_n(D-2n)}{(D-2n)!}
\eps_{a_1a_2\cdots a_{D-1}a} R^{a_1a_2}\wedge 
\cdots R^{a_{2n-1}a_{2n}}
\wedge e^{a_{2n+1}}\wedge \cdots 
\wedge e^{a_{D-1}}=0 
\,.
\label{eqofm}
\ee

We can write the curvature two--form as
\be
R^{ab}=C^{ab}+\frac1{D-2}(e^a\wedge K^b-e^b\wedge K^a)
\label{eq:R}
\ee
where $C^{ab}$ is a two--form composed of the Weyl tensor 
$C_{\mu\nu\rho\si}$ while $K^a$ is a one--form defined (for invertible
vielbeins $e_\mu^a$) as 
$K^a=K_{\mu\nu}e^{\mu a}dx^{\nu}=K^a{}_b e^b$ with
\be
K_{\mu\nu}\equiv R_{\mu\nu}- \frac1{2(D-1)}g_{\mu\nu}R
\,. 
\ee

In this paper we consider the background metrics for which the Weyl
tensor vanishes. In this case the curvature two--form (\ref{eq:R}) 
can be expressed in terms of the Ricci tensor and the curvature scalar
only.

Multiplying (\ref{eqofm}) by $e^b$ we get
\bea
0&=&
e\sum_n\ka_n
\frac{2^{n}(D-n-1)!}{(D-2)^n(D-2n-1)!} 
\left(\de_{a_2}^{c_2}\de_{a_4}^{c_4}
\cdots\de_{a_{2n}}^{c_{2n}}\de_a^b
\pm{\rm perm.}\right)
K^{a_2}_{c_2}K^{a_4}_{c_4}\cdots K^{a_{2n}}_{c_{2n}}
\nn\\
&=&
e\sum_n\ka_n(-1)^{n}
\frac{2^{n}n!(D-n-1)!}{(D-2)^n(D-2n-1)!}H^{(n)}_{ab}
\eea
where $H^{(n)}$ are defined by this equality and will be described in
detail below.

It is not too difficult to prove that the generating functional for
the tensors $H_{\mu\nu}^{(n)}$ is given by: 
\be
M_{\mu\nu}\left( t,K\right)
=
\det(\id - t K)\left((\id- t K)^{-1}\right)_{\mu\nu}
=
\sum_{n=0}^\infty  t^n \left(H^{(n)}\right)_{\mu\nu}
\label{eq:M}
\ee
where from now on we use the matrix notation 
($\id$ denotes $\de_\mu^\nu$ while $K$ denotes $K_\mu{}^\nu$). 

Differentiating equality $C_{\mu\nu\rho\si}=0$ over  
$x^\mu$ and using Bianchi identity we get 
\be
K_{\mu\nu;\rho}=K_{\mu\rho;\nu}
\,.
\label{KK}
\ee

By differentiating $M_{\mu\nu}\left( t,K\right)$ over $x^{\mu}$
and using (\ref{KK})
it is straightforward to prove that tensors $H^{(n)}_{\mu\nu}$ are
covariantly conserved. A more tedious calculation shows a nontrivial
fact that
(when the Weyl tensor vanishes)  
$H^{(n)}_{\mu\nu}$ are the only covariantly conserved tensors which
are algebraically composed of the Ricci tensors.

{}From the generating functional (\ref{eq:M}) 
one gets the recurrence relations for $H_{\mu\nu}^{(n)}$:
\be
H_{\mu\nu}^{(n)}=H_{\mu\rho}^{(n-1)} 
K^\rho{}_\nu-\frac1{n}g_{\mu\nu}H_{\si\rho}^{(n-1)}K^{\si\rho}
\ee
starting with $H_{\mu\nu}^{(0)}=g_{\mu\nu}$;  
the explicit formula (for $n>0$) reads
\be
H^{(n)}_{\mu\nu}
=
\!\!\!
\sum_{\{q_1,\ldots,q_n\}}
\left[\prod_{m=1}^n\frac{(-\Tr K^m)^{q_m}}{m^{q_m} q_m!}\right]
\left(K^{n-Q}\right)_{\mu\nu}
\label{eq:Hexplicit}
\ee
where $Q=\sum_{k=1}^n kq_k$ and the sum is over all sets of $n$
non--negative integers $q_k$ for which $Q\le n$.

As an example we present the first four tensors $H^{(n)}_{\mu\nu}$:
\bea
H_{\mu\nu}^{(0)}
&\!\!=&\!\!
g_{\mu\nu}
\,,\nn\\ 
H_{\mu\nu}^{(1)}
&\!\!=&\!\!
R_{\mu\nu}-\frac12 R g_{\mu\nu}
\,,\nn\\ 
H_{\mu\nu}^{(2)}
&=&
R_{\mu\rho} R^\rho{}_{\nu}-\frac{D}{2(D-1)}R R_{\mu\nu}
+
g_{\mu\nu}
\left(-\frac12 R_{\si\rho}R^{\si\rho}+\frac{D+2}{8(D-1)} R^2 \right) 
\,,
\nn\\
H_{\mu\nu}^{(3)}
&=&
R_{\mu\si}R^{\si\rho}R_{\rho\nu}
-\frac{D+1}{2(D-1)}RR_{\mu\si}R^\si{}_\nu
+R_{\mu\nu}\left(
-\frac12 R_{\rho\si}R^{\rho\si}
+\frac{D^2+3D-2}{8(D-1)^2}R^2\right)
\nn\\
&&
+g_{\mu\nu}\left(-\frac13 R_{\tau\si}R^{\si\rho}R_{\rho\tau}
+\frac{D+1}{4(D-1)}RR_{\rho\si}R^{\rho\si}
-\frac{D^2+7D+2}{48(D-1)^2}R^3\right)
\,.
\eea
$H^{(2)}_{\mu\nu}$ was used (for $D=4$) in \cite{S} to 
prove that the Einstein--Hilbert gravity with such a correction does
not have a cosmological singularity and can describe inflation.

One can see from the definition (\ref{eq:I}) that in $D$ space--time
dimensions forms $\mbox{\boldmath$I$}^{(n)}$ are nonzero only for 
$0 \le n\le [D/2]$ (therefore also $n_{max} \le[D/2]$ in
(\ref{eq:L})). 
The situation is different for the
tensors $H^{(n)}_{\mu\nu}$. Using the 
generating functional (\ref{eq:M}) or the explicit formula
(\ref{eq:Hexplicit}) it is straightforward to show that
$H^{(n)}_{\mu\nu}$ are nonzero in the larger range $0 \le n<D$. 
The difference arises because the energy--momentum tensors 
$T^{(n)}_{\mu\nu}$ derived from $\mbox{\boldmath$I$}^{(n)}$ are
proportional to $H^{(n)}_{\mu\nu}$ with coefficients vanishing for  
$n \ge D/2$:
\be
T^{(n)}_{\mu\nu}
=
\ka_n\frac{(-2)^{n-1}n!(D-n-1)!}{(D-2)^n(D-2n-1)!}
H^{(n)}_{\mu\nu}
\,.
\label{eq:T}
\ee
The case with $D=2n$ should be treated with 
care because it is related to the conformal anomaly.

Let us illustrate this point by considering
$n=2$. The Euler density in this case is the famous Gauss--Bonnet 
combination: $I^{(2)}=R_{\al\beta\rho\si}R^{\al\beta\rho\si} 
-4R_{\al\beta}R^{\al\beta}+R^2$. 
The energy--momentum tensor obtained from this density is given by 
(\ref{eqofm}):
\be
T^{(2)}_{\mu\nu}
=\ka_2\left(-4R^{\al\beta}R_{\al\mu\beta\nu}+
2R_{\mu\al\beta\ga}R_\nu{}^{\al\beta\ga}
-4R_{\mu\al}R^\al{}_\nu 
+2RR_{\mu\nu}-\frac12 g_{\mu\nu}I^{(2)}\right)
\,. 
\label{eq:T2}
\ee
Let us note that this energy--momentum tensor is covariantly
conserved even with nonvanishing Weyl tensor. 
Relation (\ref{eq:T2}) for the conformally flat metric 
gives 
\be
T^{(2)}_{\mu\nu}=-\ka_2\frac{4(D-4)(D-3)}{(D-2)^2} 
H^{(2)}_{\mu\nu} 
\,.
\ee
The coefficient in the above equality contains the factor $(D-4)$ but
it is compensated by a factor $\Ga(D-4)$ present in $\ka_2$ due to
regularization of the effective action so 
$T^{(D/2)}_{\mu\nu}$ should be taken into account.

Let us now use the above formalism to investigate the
possibility of warped metric solutions in theories with higher order
gravity terms in the lagrangian (\ref{eq:L}).

In this paper we will consider a model in $D$-dimensional
space--time with a $(D-2)$--brane (a domain wall). 
Its action is the sum of the bulk and brane contributions:
\bea
S
&=&
S_{\rm bulk}+S_{\rm brane}
\,,
\nn\\
S_{\rm bulk}
&=&
- \int d^Dx \sqrt{-g} \sum_{n=0}^{n_{max}} \ka_n I^{(n)}
\,,
\label{eq:S}
\\
S_{\rm brane}
&=&
- \int d^{D-1}x \sqrt{-\tilde g} \left(\la + \ldots \right)
\nn
\,.
\eea
The metric on the brane is given by 
${\tilde g}_{\mu\nu}(x^\rho)=g_{\mu\nu}(x^\rho,y=0)$ 
where $y=x^D$ and from now on    
$\mu,\nu,\ldots=1,\ldots,D-1$ while $M,N,\ldots=1,\ldots,D$. 
In the brane action we write explicitly only the most important term
-- the brane cosmological constant. The bulk gravitational 
interactions are described by the sum of terms $I^{(n)}$ 
defined in (\ref{eq:I}). 
The first two 
terms are known from conventional gravity. The one with $n=1$ 
is the usual Hilbert--Einstein term, $I^{(1)}=R$, and its 
coefficient, $\ka_1$, depends on the fields normalization. We 
will use normalization for which $\ka_1=1$ (the other frequent choice
is $(2\ka^2)^{-1}$). The term with $n=0$ corresponds to  
the cosmological constant: $I^{(0)}=1$, $\ka_0=\La$. The 
maximal number of the higher order terms is $n_{max} \le [D/2]$ 
as previously discussed.

We obtain the equations of motion by differentiating
the action (\ref{eq:S}) with respect to $g^{MN}$:
\be
R_{MN}-\frac{1}{2}Rg_{MN}
=
\frac{1}{2}\La g_{MN}
+ \frac{1}{2}\la {\tilde g}_{\mu\nu} \de^\mu_M \de^\nu_N \de(y)
- \sum_{n=2}^{n_{max}} T^{(n)}_{MN}
\,.
\label{eq:EoM}
\ee
We see that $T^{(n)}_{MN}$ given by (\ref{eq:T}), for $n \ge 2$ 
can be treated as contributions
to the energy--momentum tensor coming from higher order
interactions present
in the action (\ref{eq:S}).

Let us now look for the domain wall
solutions in $D$--dimensional space--time which are flat from the 
($D-1$)--dimensional point of view. 
The metric in this case can be written in the form
\be
ds^2=e^{-2f(y)} \eta_{\mu\nu} dx^\mu dx^\nu + dy^2
\,.
\label{eq:ds2}
\ee
Using the generating functional (\ref{eq:M})
we can find $H^{(n)}_{MN}$ for this metric:
\be
H^{(n)}_{MN}
=
g_{MN}\left(\frac{D-2}{2}\right)^n
\left(\!\!
\ba{c}{D-1}\\{n}\ea
\!\!\right)
\left(\frac{\pa f}{\pa y}\right)^{2n-2}
\left[
\left(\frac{\pa f}{\pa y}\right)^2
-
 \left(1-\de^M_D\right)
\frac{2n}{D-1}
\left(\frac{\pa^2 f}{\pa y^2}\right)
\right]
\!.
\label{eq:Hdw}
\ee
This explicit formula shows that all $H^{(n)}_{MN}$
vanish for $n \ge D$ and that the second derivative of $f(y)$ is
absent
in $H^{(n)}_{DD}$ and
appears only linearly in $H^{(n)}_{\mu\nu}$. 
This is quite surprising for a theory with higher
orders of the curvature tensor in the action and, as we will see
shortly, is crucial for the existence of the domain wall solutions.
This property follows from the unique structure of the Euler densities
(\ref{eq:I}) and the fact that the second derivative of $f(y)$ appears
only in $K_{DD}$ and is absent in $K_{\mu\nu}$. Although higher powers
of ${\pa^2 f(y)}/{\pa y^2}$ are present in separate terms on the
r.h.s. of eq.\ (\ref{eq:Hexplicit}) they exactly cancel in the full
sum.

Using tensors (\ref{eq:Hdw}) we find that
the equations of motion (\ref{eq:EoM}) for the domain wall metric
(\ref{eq:ds2}) reduce to just two conditions:
\bea
\sum_{n=1}^{n_{max}}
p_n
\left(\frac{\pa f}{\pa y}\right)^{2n}
&=&
\frac{1}{2}\,\La
\,,
\label{eq:sol_bulk}
\\
\sum_{n=1}^{{n_{max}}}
np_n
\left(\frac{\pa f}{\pa y}\right)^{2n-2}
\frac{\pa^2 f}{\pa y^2}
&=&
-\frac{D-1}{4}\,\la\,\de(y)
\label{eq:sol_brane}
\eea
where the coefficients $p_n$ are equal
\be
p_n
=
\ka_n \frac{(-1)^{n-1}(D-1)!}{2(D-1-2n)!}
\,.
\ee
The absence of the third or higher derivatives of $f(y)$ and/or higher
powers of the second derivative ${\pa^2 f(y)}/{\pa y^2}$ in
(\ref{eq:Hdw}) allows for the solution of (\ref{eq:EoM}) 
of the same type as in the standard gravity:
\be
f(y)=\si|\,y|
\ee
where the parameter $\si$ in this case must satisfy two
algebraic 
conditions coming from (\ref{eq:sol_bulk},\ref{eq:sol_brane}): 
\bea
\sum_{n=1}^{{n_{max}}} p_n \si^{2n}
&=&
\frac{1}{2}\,\La
\,,
\label{eq:si_bulk}
\\
\sum_{n=1}^{{n_{max}}} \frac{n}{2n-1} p_n \si^{2n-1}
&=&
-\frac{D-1}{8}\,\la
\,.
\label{eq:si_brane}
\eea
One of these equations can be used to determine the warp
factor coefficient $\si$ in terms of the parameters $p_n$ and
one of the cosmological constants (bulk, $\La$, or brane, 
$\la$). The second equation is then the (fine tuning) 
condition for the other cosmological constant.

Recently there has been a lot of interest in domain wall
solutions in theories with the space--time dimension equal to 5.
The Randall--Sundrum scenario
{\cite{RS2} corresponds to $D=5$ and 
only two terms with $n=0,1$ ($n_{max}=1$) in the lagrangian.
In such a case the square of the warp factor exponent,
$\si^2$, is just proportional to the bulk cosmological constant
$\La$ and the brane cosmological constant $\la$ must be
appropriately fine tuned: $\si\la=-\La$.

However, the equations (\ref{eq:si_bulk},\ref{eq:si_brane}) 
have other very interesting solutions when the 
higher order terms are present (i.e. $n_{max} \ge 2$). 
The square of the warp factor exponent 
$\si^2$ is no longer just linearly proportional to 
$\La$. It depends also on the other lagrangian parameters 
$\ka_n$. In general there are up to $n_{max}$ possible values of 
$\si^2$ for a given value of $\La$. This is true also for 
$\La=0$. Thus, it is possible to have nontrivial warped metric
solutions with vanishing bulk cosmological constant  
for any $n_{max} \ge 2$. There is one such solution for each positive 
zero of the polynomial $P(x)=\sum_{n=1}^{{n_{max}}} p_n x^n$. 
It is also possible to solve 
eqs.\ (\ref{eq:si_bulk},\ref{eq:si_brane}) with $\lambda=0$.
The domain wall is then supported only by higher order corrections to 
the action and (in distinction to the Einstein--Hilbert gravity) 
there is no need for conventional sources on the
brane.

Even more interesting situation emerges when
the zero of $P(x)$ satisfies simultaneously eq.\ (\ref{eq:si_brane}).
Then, there exists such a warp factor $\si$ that 
eq.\ (\ref{eq:si_bulk}) is satisfied for $\La=0$ 
and eq.\ (\ref{eq:si_brane}) is satisfied for $\la=0$.
In such a case, there is a domain wall solution which is flat from
the   
($D-1$)--dimensional point of view with both 
cosmological constants vanishing. This, however, can happen 
only for $n_{max} > 2$ (i.e. $D \ge 6$) 
and for some very specific values of the 
lagrangian parameters $\ka_n$. The fine tuning of parameters $\ka_n$
replaces the  fine tuning of the brane tension $\la$ used in the
original Randall--Sundrum scenario. Therefore, it is a different
formulation of the cosmological constant problem.

The domain wall solutions with the first correction (Gauss--Bonnet
term) were discussed in different contexts in the literature. 
The 5--dimensional Gauss--Bonnet gravity
with dilaton was discussed in detail in \cite{MR}.
The brane solutions with vanishing bulk cosmological constant (but
with non-vanishing brane cosmological constant) were
discussed in gravity coupled to the dilaton
(see for example \cite{KSS,LZ} where the first reference does not
discuss
the Gauss--Bonnet 
term). The mechanism of ``self--tuning'' presented in these papers was
shown later \cite{FLLN} to be another form of fine tuning.
The possibility of brane type solutions with 
vanishing cosmological constants was also disscussed in \cite{AMO}.
The idea of employing quadratic corrections as
boundary counterterms to the AdS action was used fr example in\
\cite{KLS,HHR}.  
The brane configurations in the context of AdS/CFT correspondence were
considered for example in refs.\ \cite{HHR,NO} and in the cosmological
context in \cite{KKLN}.

In conclusion,
we considered a class of models with higher order gravity corrections 
in the form of the Euler densities with arbitrary power $n$ of the
curvature tensor in arbitrary space--time dimension $D$.  
We found a generating functional and an explicit form of the
corresponding energy--momentum tensors.
The domain wall type solutions were constructed for 
the considered class of models. Such solutions are possible 
despite the presence of higher powers of the curvature
tensor because the singularity structure at the wall 
is of the same type as in the standard gravity. 
The number of possible domain wall solutions increases with the
order of included corrections and the cosmological constant is
no longer crucial for their existence (as it was the case for models 
with $n \le 2$ previously analysed in the literature).
We showed for example that the Randall--Sundrum type scenario can be
realized without the bulk and/or brane cosmological constant.

This work was partially supported by the Polish KBN grants 
5 P03B 150 20 and 2 P03B 052 16.

\end{document}